\documentclass[preprint,prb]{revtex4-1}

\usepackage{upgreek}
\usepackage[latin1]{inputenc}
\usepackage{subsupscripts}
\usepackage{subscript}

\usepackage{graphicx}
\usepackage{dcolumn}
\usepackage{bm}
\usepackage{color}

\begin{document}

\title{Nonlinear spectroscopy of exciton-polaritons in a GaAs-based microcavity}
\author{Johannes Schmutzler$^1$, Marc A{\ss}mann$^1$, Thomas Czerniuk$^1$, Martin Kamp$^2$, Christian Schneider$^2$, Sven H\"ofling$^{2,3}$ and Manfred Bayer$^{1,4}$}
\address{$^1$Experimentelle Physik 2, Technische Universit\"at Dortmund, \mbox{D-44221 Dortmund, Germany}}
\address{$^2$Technische Physik, Physikalisches Institut, Wilhelm Conrad R\"ontgen Research Center for Complex Material Systems,
Universit\"at W\"urzburg, D-97074 W\"urzburg, Germany}
\address{$^3$SUPA, School of Physics and Astronomy, University of St Andrews, St Andrews, KY16 9SS, United Kingdom}
\affiliation{$^4$A. F. Ioffe Physical-Technical Institute, Russian Academy of Sciences, St Petersburg 194021, Russia}

\date{5 August 2014}

\begin{abstract}
We present a systematic investigation of two-photon excitation processes in a GaAs-based microcavity in the strong-coupling regime. We observe second harmonic generation resonant to the upper and lower polariton level, which exhibits a strong dependence on the photonic fraction of the corresponding polariton. In addition we have performed two-photon excitation  spectroscopy to identify $2p$ exciton states which are crucial for the operation as a terahertz lasing device, which was suggested recently  [A. V.
Kavokin et al., Phys. Rev. Lett. \textbf{108}, 197401 (2012)]. However, no distinct signatures of a $2p$ exciton state could be identified, which indicates a low two-photon pumping efficiency.

\end{abstract}



\maketitle
\section{Introduction}
Exciton-polaritons in semiconductor microcavities have gained increasing attention in the last 20 years since they are not only of interest for fundamental research as they allow for the observation of Bose-Einstein condensation (BEC) in semiconductors,\cite{Kasprzak2006,Balili2007} but also offer the potential for a wide range of applications. Especially the implementation of  exciton-polaritons as an active material for a new class of semiconductor laser devices, the so-called polariton-laser,\cite{imamoglu1996,Kammann2012,Bajoni2012} and as building blocks for integrated photonic circuits\cite{Liew2008,Ortega2013,Sturm2014} is promising. Recently, there was a proposal for a vertical cavity surface emitting terahertz (THz) laser based on the stimulated THz transition between the dipole-forbidden $2p$ exciton state and the lower exciton-polariton in a microcavity.\cite{Kavokin2012} Thereby, a two-photon pumping process of the $2p$ state is necessary due to selection rules.

While THz spectroscopy is an appealing tool for a wide range of applications such as the investigation of biomolecules \cite{Fischer2005}, material evaluation \cite{Nagai2004} and for security issues, e. g. tracing of illegal drugs \cite{Kawase2003}, (a comprehensive overview of applications of THz spectroscopy can be found in Ref.~\onlinecite{Tonouchi2007}) the number of available laser sources in this spectral region is rather limited.
Furthermore, all coherent THz sources available so far suffer from certain drawbacks, as they are either bulky and expensive (e. g. free electron lasers), exhibit low efficiencies [e. g. THz radiation generated by optical mixing techniques \cite{Brown1995} and photocarrier acceleration in photoconducting antennas \cite{Cai1997}], or can only be operated below room temperature and require a complicated design (quantum cascade lasers \cite{Williams2007}). Therefore, the identification of new reliable THz sources is currently a dynamic field of research.\cite{Mittleman2013}

So far, there are encouraging proof of principle studies for the feasibility of polariton based lighting devices, as electrically pumped polariton lasers \cite{Schneider2013,Bhattacharya2013,Bhattacharya2014} as well as polariton light-emitting diodes have been demonstrated \cite{Tsintzos2008}, and also for the operability of logic gates and transistor switches based on exciton-polaritons.\cite{Ballarini2013,Nguyen2013,Gao2012} Concerning applications operating in the THz frequency range, oscillations of the polariton population due to the dynamic Stark effect \cite{Hayat2012,Cancellieri2014} and relaxation oscillations \cite{DeGiorgi2014} have been observed, but  unambiguous evidence for  THz-lasing operation is still lacking. For an evaluation of the feasibility of the proposal of Ref.~\onlinecite{Kavokin2012} a careful investigation of the occurring two-photon processes in a semiconductor microcavity is necessary. Recently, two-photon excitation of polaritons in a GaAs-based microcavity system was reported.\cite{Lemenager2014} In this study a femtosecond-pulsed laser with a spectral width of $13~\mbox{meV}$, more than two times larger than the Rabi splitting in the investigated sample, was used, which did not allow for energy-resolved two-photon excitation spectroscopy (TPE spectroscopy). The strong emission of the upper (UP) and lower polariton (LP) state at resonant excitation was interpreted in terms of $2p$ exciton injection followed by stimulated THz-emission into the LP state. However, the identification of a two-photon absorption (TPA) process arising from the $2p$ exciton state and the distinction from second harmonic generation (SHG) processes at the LP and UP state energies is challenging in this study due to the limited spectral resolution of the used femtosecond-pulsed laser system. 

The scope of our paper lies in a systematic investigation of the occurring two-photon excitation processes in a GaAs-based microcavity. Here, we use a tunable nanosecond-pulsed OPO with a spectral width of $0.3~\mbox{meV}$ allowing for energy-resolved TPE and SHG spectroscopy. Thereby we are able to differentiate between TPA and SHG processes. While we do not observe distinct signatures of a $2p$ exciton in the TPE spectra, we see strong SHG emission from the LP and UP instead, which is dependent on the photonic fraction of the corresponding polariton.

This manuscript is structured as follows: In Sec.~II the investigated sample as well as the experimental techniques used are described. This is followed by a presentation of our experimental results in Sec.~III, which covers TPE as well as SHG spectroscopy experiments. Finally, a conclusion and an outlook for further experiments is given in Sec. IV.

\section{Experimental details}
\label{Experimental details}
We investigate a GaAs-based $\lambda/2$ microcavity with a quality factor of ca. $1800$. The design of the sample is as follows:
Three stacks of four GaAs quantum wells are placed in the three central antinodes of the electric field confined by two distributed Bragg reflector (DBR) structures in a $\uplambda$/2 cavity. The upper (lower) DBR structure consists of 16 (20) alternating layers of Al\textsubscript{0.2}Ga\textsubscript{0.8}As and AlAs. The interaction of the cavity field with the exciton resonance of the 12 contained GaAs quantum wells leads to a Rabi splitting of ca. $14~\mbox{meV}$.

The sample is mounted in a helium-flow cryostat, which allows for experiments at temperatures as low as $10~\mbox{K}$. For two-photon excitation a
nanosecond-pulsed optical-parametric oscillator (OPO) pumped by the third harmonic of a Nd:YAG laser is used. The repetition rate of the laser system is $10~\mbox{Hz}$. For TPE spectroscopy the elliptically shaped laser beam is focused under 45° degrees of incidence onto the sample with main axes of $x=100~\upmu\mbox{m}$ and $y=500~\upmu\mbox{m}$, respectively, which is applied for the experiments presented in Sec.~\ref{TPE spectroscopy}. For SHG spectroscopy excitation under normal incidence is chosen for the sake of wavevector conservation. Thereby the  main axes of the elliptically shaped spot are $x=20~\upmu\mbox{m}$ and $y=100~\upmu\mbox{m}$, respectively, which is applied for the experiments presented in Sec.~\ref{SHG spectroscopy}.

The investigated sample exhibits a gradient of the exciton-cavity detuning of $2.5~\mbox{meV}/\mbox{mm}$ concerning the x-axis of the excitation laser spot and is about constant in y-direction. Therefore, even for the larger excitation laser spot, the detuning of the probed polaritons is about constant. 

TPE spectroscopy is used to investigate nonresonant TPA processes. Thereby the detection energy remains fixed in an energy range of $3.3~\mbox{meV}$ centered around the LP energy, which corresponds to an integration of an ensemble of LP states covering an in-plane momentum range of $k_{||}=\pm1.8~\upmu\mbox{m}^{-1}$. Here, the excitation energy of the OPO is tuned in a range of $792-846~\mbox{meV}$.

Furthermore, we use SHG spectroscopy as a complementary experimental technique, in which only resonant two-photon processes are considered.
Here, the detection energy is tuned in combination with the wavelength of the OPO and corresponds always to the two-photon energy of the OPO.
The combination of both experimental techniques using a nanosecond-pulsed OPO system with a rather narrow linewidth of $0.3~\mbox{meV}$ compared to femtosecond-pulsed laser systems allows for a clear distinction between nonresonant TPA processes and coherent SHG.

In all experiments the emission from the sample is collected under normal incidence using a microscope objective (numerical aperture $0.26$) and  detected spectrally resolved by a thermoelectric-cooled CCD-camera behind a monochromator.

\section{Results and discussion}

This section is split into two parts. In Sec.~\ref{TPE spectroscopy} nonresonant TPA processes are studied and the feasibility of an efficient population of the $2p$ exciton is evaluated, whereas in Sec.~\ref{SHG spectroscopy} coherent SHG processes in GaAs-based microcavities are revealed.

\subsection{TPE spectroscopy}
\label{TPE spectroscopy}

\begin{figure}[!htb]
\centering
\includegraphics[width=0.85\linewidth]{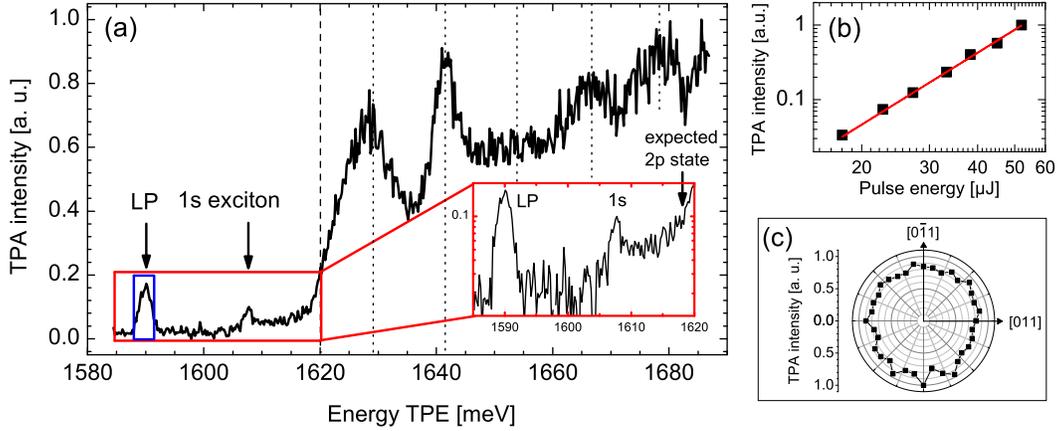}
\caption{(Color online) (a) TPE spectrum for a detuning $\delta=E_c-E_x=-16.8~\mbox{meV}$, where $E_c$ ($E_x$) denotes the energy
of the cavity (exciton). The pulse energy of the OPO is kept constant at ca. $50~\upmu\mbox{J}$. The onset of the lowest intersubband transition is indicated by the black dashed line. The oscillations superimposed on the increasing signal intensity above $1620~\mbox{meV}$ due to the DBR are marked by black dotted lines. The detection energy range is indicated by the blue rectangle. Inset: Close-up of the TPE spectrum below the intersubband transition. Above $1605~\mbox{meV}$ in energy an increase of the TPE signal can be observed. (b) Dependence of signal intensity of the LP on the pulse energy of the OPO for a two-photon energy of $1653~\mbox{meV}$. Black squares, data; red line, power-law fit $y=a\cdot x^p$ with $p=3.20\pm0.15$. (c) Polarization dependence of the TPA intensity for an incoming polarization of the OPO in $[011]$ direction. Signal intensity is proportional to the radial distance from the center. Two-photon energy lies at $1653~\mbox{meV}$, the pulse energy is $42~\upmu\mbox{J}$.}
\label{Fig1}
\end{figure}

\begin{figure}[!htb]
\centering
\includegraphics[width=0.65\linewidth]{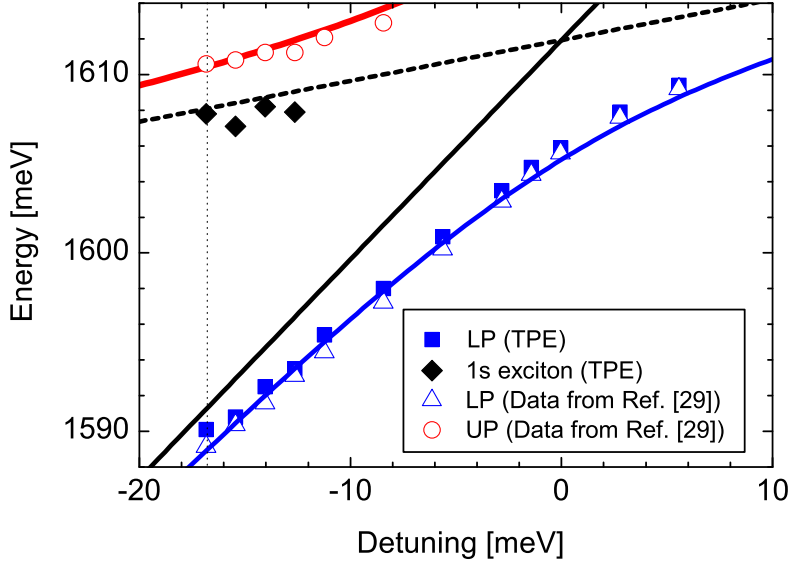}
\caption{(Color online) Dependence of the observed resonance energies below the sub-bandgap on the exciton-cavity detuning. Solid symbols are taken from TPE spectra. Open symbols are taken from Ref.~\onlinecite{Schmutzler2013a}. Black solid line (dashed line) is the calculated cavity ($1s$ exciton) energy according to Ref.~\onlinecite{Schmutzler2013a}. TPE data indicated by the vertical dotted line are extracted from the TPE spectrum presented in figure~\ref{Fig1}.}
\label{Fig2}
\end{figure}

In figure~\ref{Fig1} a typical TPE spectrum at a negative detuning of $\delta=-16.8~\mbox{meV}$ is presented.
The strong rise in signal intensity above $1620~\mbox{meV}$ is attributed to the intersubband transition between the highest heavy hole subband and the lowest conduction subband. The oscillations superimposed on the increasing signal intensity above $1620~\mbox{meV}$ with a periodicity of roughly $12~\mbox{meV}$ (indicated by black dotted lines in figure~\ref{Fig1}) are caused by a modulation of the reflectivity in the infrared spectral range due to the DBR. The comparison with data from previous work \cite{Schmutzler2013a} (figure~\ref{Fig2}) allows for a clear identification of the two resonances below $1620~\mbox{meV}$ (figure~\ref{Fig1}) as $1s$ exciton and LP, respectively. The LP energies observed here, are slightly higher in energy compared to Ref.~\cite{Schmutzler2013a} due to the integration over a broad distribution of LP states with different in-plane momenta, whereas the data of Ref.~\onlinecite{Schmutzler2013a} are related to the LP state at $k_{||}=0$. We do not observe any resonance corresponding to the UP energy in the TPE spectra (figures~\ref{Fig1} and \ref{Fig2}). Note, that the radiative transition between the UP and LP state is forbidden due to parity conservation, which can only be overcome for the case of the hybridization of the UP with an exciton exhibiting a different parity.\cite{Kavokin2010} However, this requires a careful design of the sample and is not a generic effect. Further, for a phonon-assisted relaxation process, the emission of  acoustic phonons is required, as the energy splitting between the UP and LP is roughly a factor of two smaller than the longitudinal optical phonon energy. Therefore, we believe, that SHG is the dominating process compared to TPE of the LP via the UP. 

Scattering from the $1s$ exciton to LP states is only observed at large negative detunings, corresponding to a photonic fraction of the LP ground state larger than $80~\%$. The close-up of the TPE spectrum [inset of figure~\ref{Fig1} (a)] reveals an onset of the TPA process roughly at $1605~\mbox{meV}$. Beside the $1s$ exciton, no further exciton resonances can be observed in the TPE spectrum.

\begin{figure*}[!htb]
\centering
\includegraphics[width=0.90\linewidth]{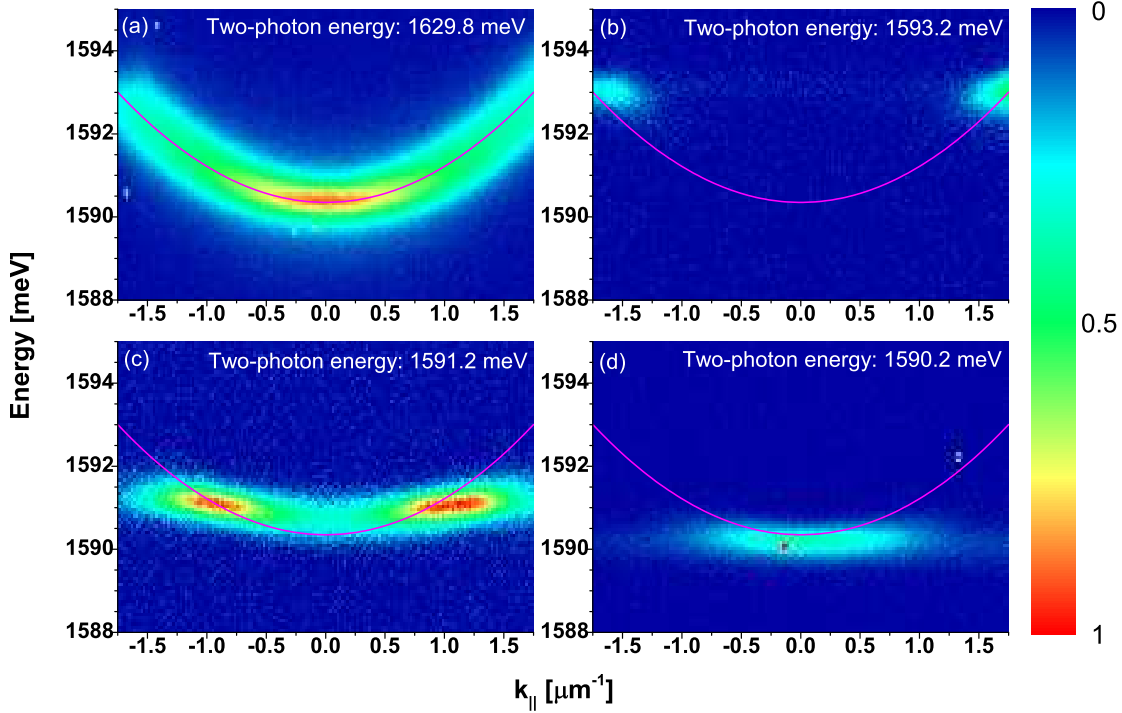}
\caption{(Color online) Far field emission for excitation with different two-photon energies at $\delta=-15.4~\mbox{meV}$. The excitation pulse energy is ca. $50~\upmu\mbox{J}$. (a) Emission of polaritons exhibiting a broad distribution of in-plane momenta can be observed under nonresonant excitation. (b)-(d) For the case of resonant excitation only emission of polaritons corresponding to the two-photon energy of the OPO can be seen. The solid line corresponds to the calculated LP dispersion.}
\label{Fig3}
\end{figure*}

\begin{figure}[!t]
\centering
\includegraphics[width=\linewidth]{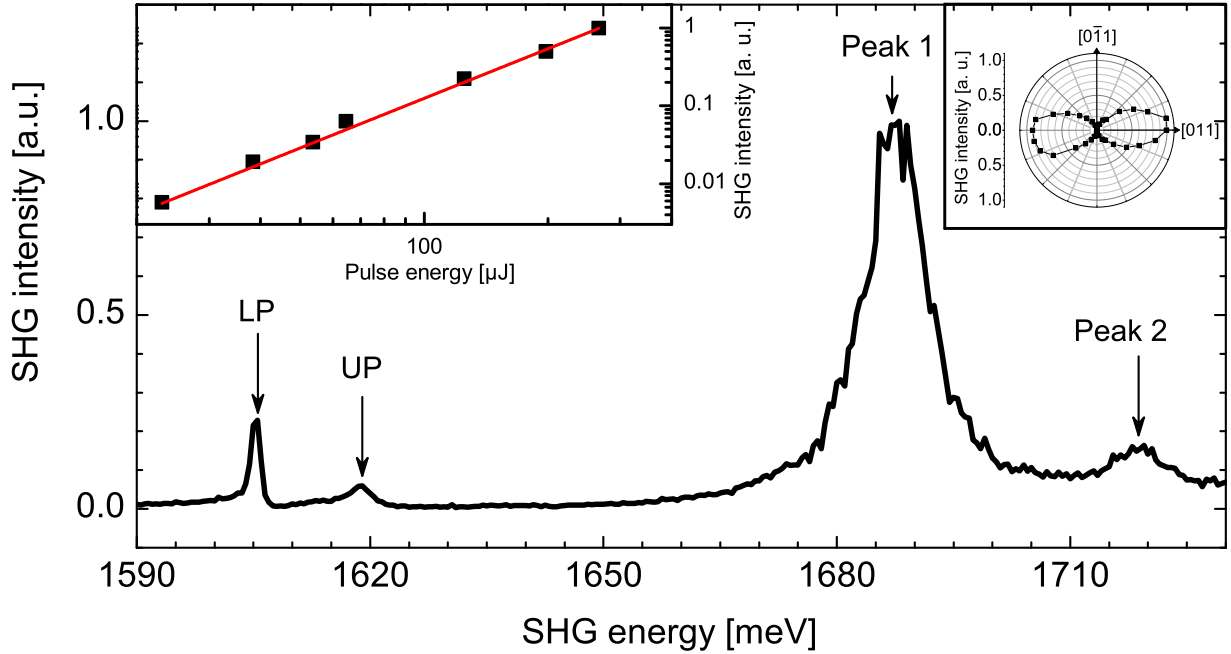}
\caption{(Color online) SHG spectrum for $\delta=-0.7~\mbox{meV}$ and an excitation pulse energy of $20~\upmu\mbox{J}$. LP and UP resonances can clearly be identified. Left inset: Dependence of signal intensity of the LP resonance on the pulse energy of the OPO. Black squares, data; red line, power law fit $y=a\cdot x^p$ with $p=2.11\pm0.09$. Right inset: Polarization dependence of the SHG emission at the LP energy for an incoming polarization of the OPO in $[011]$ direction. Signal intensity is proportional to the radial distance from the center. The excitation pulse energy is $20~\upmu\mbox{J}$.}
\label{Fig4}
\end{figure}
\noindent
The $2p$ exciton is expected to be roughly $10~\mbox{meV}$ higher in energy compared to the $1s$ exciton in GaAs quantum wells\cite{Nithisoontorn1989}, which is only $2~\mbox{meV}$ separated in energy from the intersubband transition. A clear identification of the $2p$ exciton is therefore challenging.

This finding coincides with previous studies: 
While clear evidence for P-states in TPE experiments \cite{Catalano1989,Catalano1990} between the $N=2,3,4$ subbands of GaAs quantum wells was reported, only unpronounced steps in photoconductivity measurements \cite{Nithisoontorn1989} or even only a broad tail \cite{Tai1989} in TPE signal intensity below the sub-bandgap were observed as signatures of the $2p$ exciton of the lowest $N=1$ intersubband transition in GaAs-based quantum wells. 

However, the absence of a pronounced $2p$ exciton resonance might indicate a low  two-photon pumping efficiency of the LP via this channel. The detection of THz radiation is a challenging task, typically bolometers at cryogenic temperatures are used.\cite{Brown1995,Chen1997} Therefore, we suggest a precise measurement of the LP and $2p$ exciton energy levels as a mandatory preliminary experiment, which determines the exact THz frequency to be evidenced using a bolometer. 
In our opinion, materials with a higher exciton-binding energy such as ZnO or GaN are more promising for the realization of a THz laser device as proposed in Ref.~\onlinecite{Kavokin2012}, since the identification of $2p$ exciton states should be more feasible in this materials. 

In figure~\ref{Fig1} (b) the dependence of the LP emission on the pulse energy of the OPO is shown for the case of a nonresonant two-photon excitation at a two-photon energy of $1653~\mbox{meV}$. The data can be approximated well using a power-law function $y=a\cdot x^{3.2\pm0.15}$. At first sight this seems to be surprising since there is a quadratic dependence expected for a TPA process. However, since the signal  is detected at the LP, not only the creation of electron-hole pairs above the sub-bandgap contributes, but also the relaxation mechanism into the LP ground state is important. Depending on the in-plane momentum of the LP states, the following processes are of relevance: (i) spontaneous acoustic-phonon scattering for polaritons with large wavevectors and (ii) polariton-polariton scattering for LPs at $k_{||}=0$.\cite{Bloch2005} The scattering probability of the first process is independent of the reservoir density, and the latter one exhibits a quadratic dependence. In combination with the quadratic TPA process this gives an overall quadratic process for the case of acoustic phonon scattering and a fourth degree polynomial dependence for the case of polariton-polariton scattering. The cubic behavior observed here demonstrates that both mechanisms are of relevance since the emission from a broad distribution of wavevectors is detected. Note, however, that the dynamic range for the choice of the pulse energy is limited to roughly $50~\upmu\mbox{J}$ due to the onset of the destruction threshold of the sample in the case of nonresonant excitation. Figure~\ref{Fig1} (c) shows the polarization dependence of the TPA intensity under nonresonant two-photon excitation. Clearly, an isotropic emission can be observed as expected for a nonresonant excitation process.

Figure~\ref{Fig3} shows the far field emission for different excitation energies. In the case of a nonresonant excitation, where the two photon-energy is larger than the sub-bandgap, a broad distribution of LPs with different wavevectors is populated [figure \ref{Fig3}(a)]. In contrast, only LPs with the corresponding two-photon energies can be observed for the case of resonant excitation [figure \ref{Fig3}(b)-(d)]. Further, there is no evidence for a relaxation towards the LP ground state at $k_{||}=0$ [figure \ref{Fig3}(b)], which indicates a SHG process.

\subsection{SHG spectroscopy}
\label{SHG spectroscopy}
To substantiate this interpretation, in addition. we have performed SHG spectroscopy.
Thereby only resonant two-photon processes are considered, as outlined in Sec.~\ref{Experimental details}. 
Figure~\ref{Fig4} shows a typical SHG spectrum. LP and UP resonances can clearly be identified, and moreover, two more resonances at $1688~\mbox{meV}$ and $1720~\mbox{meV}$ can be observed. These peaks correspond to the reflection minima of the DBR structure, that we will discuss later in detail.  The dependence of the LP SHG intensity on the pulse energy of the OPO exhibits a quadratic behavior as expected for a SHG process (left inset of figure~\ref{Fig4}) in contrast to the observed cubic behavior for the TPA process as discussed before. Furthermore, a pronounced polarization anisotropy of the emission is observed, which gives additional evidence for SHG (right inset of figure~\ref{Fig4}), whereas the polarization of the LP emission is isotropic for a nonresonant TPA process [figure~\ref{Fig1}~(c)]. 
\begin{figure}[!htb]
\centering
\includegraphics[width=0.65\linewidth]{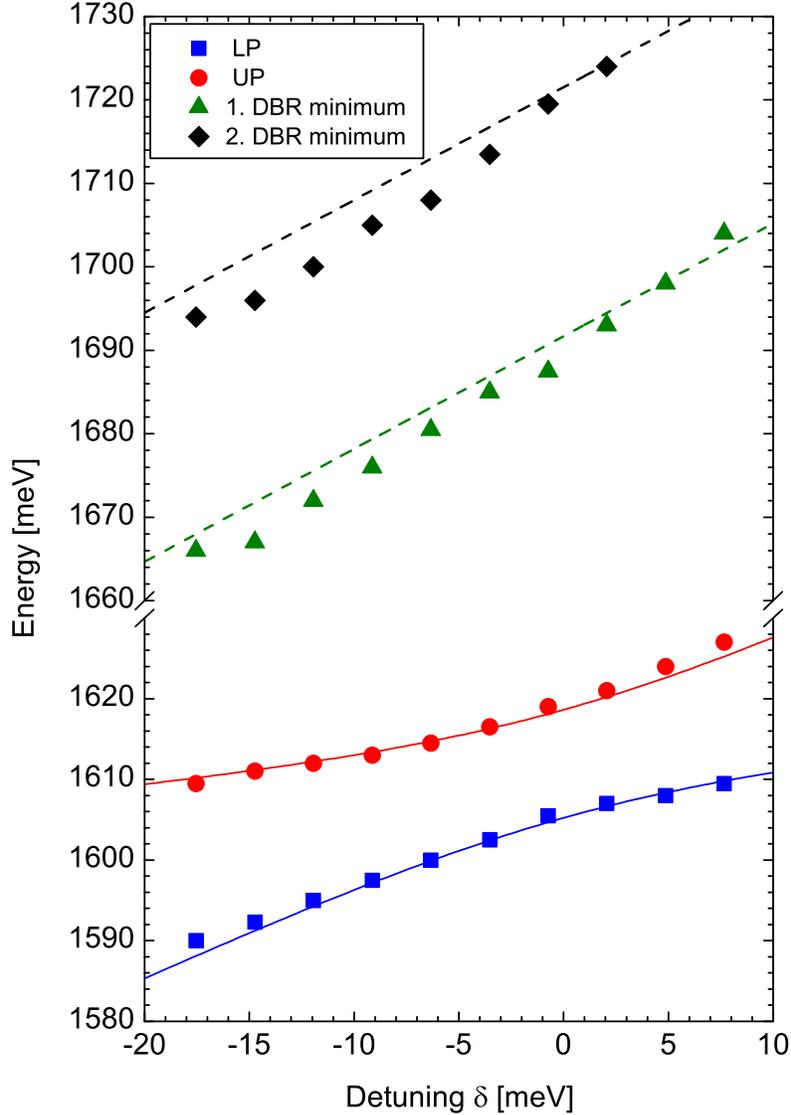}
\caption{(Color online) Dependence of the resonance energies in the SHG spectrum on detuning. Green and black dashed lines indicate the calculated energies of the first and second reflection minima of the DBR, respectively.}
\label{Fig5}
\end{figure}

\begin{figure}[!h]
\centering
\includegraphics[width=0.7\linewidth]{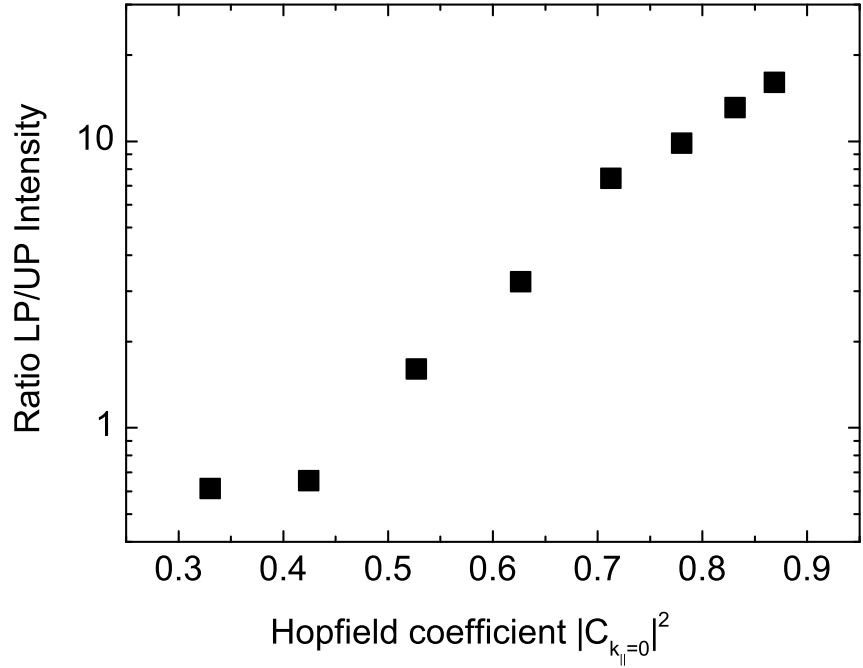}
\caption{Ratio of the SHG intensity of LP and UP depending on the Hopfield coefficient $|C_{k_{||}=0}|^2$, which determines the photonic fraction of the LP.}
\label{Fig6}
\end{figure}

SHG spectroscopy has been performed for different detunings $\delta$ (figure~\ref{Fig5}). For the LP and UP states the expected anticrossing can be seen, whereas the other two peaks exhibit a monotonous increase in energy with respect to the detuning. In addition we have performed transfer matrix calculations to determine the reflection minima of the DBR structure from the growth parameters of the investigated sample. As one can see in figure~\ref{Fig5}, the observed peaks are in good agreement with the expected energies of the reflection minima of the wedge shaped DBR. We therefore attribute these peaks to the first and second reflection minima of the DBR, respectively. 

Furthermore, we have analyzed the ratio of the intensity of the LP resonance and the UP resonance in the SHG spectra (figure \ref{Fig6}). Whereas SHG of the UP dominates for a photonic fraction of the LP of less than $50~\%$, which corresponds to a photonic fraction of the UP larger than $50~\%$, SHG of the LP is more than one order of magnitude stronger for the case of a photonic fraction of $90~\%$ of the LP. This reveals a strong dependence of the SHG efficency on the photonic fraction of the polaritons.

\section{Conclusion and outlook}
In conclusion, we have presented a systematic study of the occurring two-photon processes in a GaAs-based microcavity in the strong coupling regime. Pronounced SHG from LP, UP and reflectivity minima of the DBR have been observed, at which the photonic fraction of the polaritons is crucial for the SHG efficiency. However, clear evidence for an efficient population of the $2p$ exciton by a two-photon process is not observed by TPE spectroscopy, which is a requirement for a THz-lasing device based on microcavity polaritons.\cite{Kavokin2012} In our opinion, wide bandgap semiconductors such as GaN and ZnO, exhibiting larger exciton binding energies, are more appealing to realize such a device, as they should allow for a more straightforward identification and spectral selection of the $2p$ excitons and therefore a facilitated detection of THz radiation.

We have clearly demonstrated that SHG is a pronounced mechanism occurring in GaAs-based microcavities. We believe, that further studies investigating two-photon pumping processes of polaritons in GaAs-based microcavities should always rule out SHG to get unambiguous evidence for a TPA process.

For further experiments, a detailed study of the polarization anisotropy of the SHG at high magnetic fields should reveal the selection rules for SHG in  GaAs-based microcavities as it was demonstrated before for GaAs bulk material.\cite{Pavlov2005}

\section*{Acknowledgment}
The Dortmund group acknowledges support from the Deutsche Forschungsgemeinschaft (Grants No.~1549/18-1 and No.~1549/19-1). We are grateful to David Brunne for fruitful discussions. M.B. acknowledges support from the Russian Ministry of Science and Education (Contract No.~14.Z50.31.0021).


\begin{thebibliography}{37}%
\makeatletter
\providecommand \@ifxundefined [1]{%
 \@ifx{#1\undefined}
}%
\providecommand \@ifnum [1]{%
 \ifnum #1\expandafter \@firstoftwo
 \else \expandafter \@secondoftwo
 \fi
}%
\providecommand \@ifx [1]{%
 \ifx #1\expandafter \@firstoftwo
 \else \expandafter \@secondoftwo
 \fi
}%
\providecommand \natexlab [1]{#1}%
\providecommand \enquote  [1]{``#1''}%
\providecommand \bibnamefont  [1]{#1}%
\providecommand \bibfnamefont [1]{#1}%
\providecommand \citenamefont [1]{#1}%
\providecommand \href@noop [0]{\@secondoftwo}%
\providecommand \href [0]{\begingroup \@sanitize@url \@href}%
\providecommand \@href[1]{\@@startlink{#1}\@@href}%
\providecommand \@@href[1]{\endgroup#1\@@endlink}%
\providecommand \@sanitize@url [0]{\catcode `\\12\catcode `\$12\catcode
  `\&12\catcode `\#12\catcode `\^12\catcode `\_12\catcode `\%12\relax}%
\providecommand \@@startlink[1]{}%
\providecommand \@@endlink[0]{}%
\providecommand \url  [0]{\begingroup\@sanitize@url \@url }%
\providecommand \@url [1]{\endgroup\@href {#1}{\urlprefix }}%
\providecommand \urlprefix  [0]{URL }%
\providecommand \Eprint [0]{\href }%
\providecommand \doibase [0]{http://dx.doi.org/}%
\providecommand \selectlanguage [0]{\@gobble}%
\providecommand \bibinfo  [0]{\@secondoftwo}%
\providecommand \bibfield  [0]{\@secondoftwo}%
\providecommand \translation [1]{[#1]}%
\providecommand \BibitemOpen [0]{}%
\providecommand \bibitemStop [0]{}%
\providecommand \bibitemNoStop [0]{.\EOS\space}%
\providecommand \EOS [0]{\spacefactor3000\relax}%
\providecommand \BibitemShut  [1]{\csname bibitem#1\endcsname}%
\let\auto@bib@innerbib\@empty
\bibitem [{\citenamefont {Kasprzak}\ \emph {et~al.}(2006)\citenamefont
  {Kasprzak}, \citenamefont {Richard}, \citenamefont {Kundermann},
  \citenamefont {Baas}, \citenamefont {Jeambrun}, \citenamefont {Keeling},
  \citenamefont {Marchetti}, \citenamefont {Szymanska}, \citenamefont {Andre},
  \citenamefont {Staehli}, \citenamefont {Savona}, \citenamefont {Littlewood},
  \citenamefont {Deveaud},\ and\ \citenamefont {Dang}}]{Kasprzak2006}%
  \BibitemOpen
  \bibfield  {author} {\bibinfo {author} {\bibfnamefont {J.}~\bibnamefont
  {Kasprzak}}, \bibinfo {author} {\bibfnamefont {M.}~\bibnamefont {Richard}},
  \bibinfo {author} {\bibfnamefont {S.}~\bibnamefont {Kundermann}}, \bibinfo
  {author} {\bibfnamefont {A.}~\bibnamefont {Baas}}, \bibinfo {author}
  {\bibfnamefont {P.}~\bibnamefont {Jeambrun}}, \bibinfo {author}
  {\bibfnamefont {J.~M.~J.}\ \bibnamefont {Keeling}}, \bibinfo {author}
  {\bibfnamefont {F.~M.}\ \bibnamefont {Marchetti}}, \bibinfo {author}
  {\bibfnamefont {M.~H.}\ \bibnamefont {Szymanska}}, \bibinfo {author}
  {\bibfnamefont {R.}~\bibnamefont {Andre}}, \bibinfo {author} {\bibfnamefont
  {J.~L.}\ \bibnamefont {Staehli}}, \bibinfo {author} {\bibfnamefont
  {V.}~\bibnamefont {Savona}}, \bibinfo {author} {\bibfnamefont {P.~B.}\
  \bibnamefont {Littlewood}}, \bibinfo {author} {\bibfnamefont
  {B.}~\bibnamefont {Deveaud}}, \ and\ \bibinfo {author} {\bibfnamefont
  {L.~S.}\ \bibnamefont {Dang}},\ }\href
  {http://dx.doi.org/10.1038/nature05131} {\bibfield  {journal} {\bibinfo
  {journal} {Nature}\ }\textbf {\bibinfo {volume} {443}},\ \bibinfo {pages}
  {409} (\bibinfo {year} {2006})}\BibitemShut {NoStop}%
\bibitem [{\citenamefont {Balili}\ \emph {et~al.}(2007)\citenamefont {Balili},
  \citenamefont {Hartwell}, \citenamefont {Snoke}, \citenamefont {Pfeiffer},\
  and\ \citenamefont {West}}]{Balili2007}%
  \BibitemOpen
  \bibfield  {author} {\bibinfo {author} {\bibfnamefont {R.}~\bibnamefont
  {Balili}}, \bibinfo {author} {\bibfnamefont {V.}~\bibnamefont {Hartwell}},
  \bibinfo {author} {\bibfnamefont {D.}~\bibnamefont {Snoke}}, \bibinfo
  {author} {\bibfnamefont {L.}~\bibnamefont {Pfeiffer}}, \ and\ \bibinfo
  {author} {\bibfnamefont {K.}~\bibnamefont {West}},\ }\href {\doibase
  10.1126/science.1140990} {\bibfield  {journal} {\bibinfo  {journal}
  {Science}\ }\textbf {\bibinfo {volume} {316}},\ \bibinfo {pages} {1007}
  (\bibinfo {year} {2007})}\BibitemShut {NoStop}%
\bibitem [{\citenamefont {Imamoglu}\ \emph {et~al.}(1996)\citenamefont
  {Imamoglu}, \citenamefont {Ram}, \citenamefont {Pau},\ and\ \citenamefont
  {Yamamoto}}]{imamoglu1996}%
  \BibitemOpen
  \bibfield  {author} {\bibinfo {author} {\bibfnamefont {A.}~\bibnamefont
  {Imamoglu}}, \bibinfo {author} {\bibfnamefont {R.~J.}\ \bibnamefont {Ram}},
  \bibinfo {author} {\bibfnamefont {S.}~\bibnamefont {Pau}}, \ and\ \bibinfo
  {author} {\bibfnamefont {Y.}~\bibnamefont {Yamamoto}},\ }\href {\doibase
  10.1103/PhysRevA.53.4250} {\bibfield  {journal} {\bibinfo  {journal} {Phys.
  Rev. A}\ }\textbf {\bibinfo {volume} {53}},\ \bibinfo {pages} {4250}
  (\bibinfo {year} {1996})}\BibitemShut {NoStop}%
\bibitem [{\citenamefont {Kammann}\ \emph {et~al.}(2012)\citenamefont
  {Kammann}, \citenamefont {Ohadi}, \citenamefont {Maragkou}, \citenamefont
  {Kavokin},\ and\ \citenamefont {Lagoudakis}}]{Kammann2012}%
  \BibitemOpen
  \bibfield  {author} {\bibinfo {author} {\bibfnamefont {E.}~\bibnamefont
  {Kammann}}, \bibinfo {author} {\bibfnamefont {H.}~\bibnamefont {Ohadi}},
  \bibinfo {author} {\bibfnamefont {M.}~\bibnamefont {Maragkou}}, \bibinfo
  {author} {\bibfnamefont {A.~V.}\ \bibnamefont {Kavokin}}, \ and\ \bibinfo
  {author} {\bibfnamefont {P.~G.}\ \bibnamefont {Lagoudakis}},\ }\href
  {http://stacks.iop.org/1367-2630/14/i=10/a=105003} {\bibfield  {journal}
  {\bibinfo  {journal} {New Journal of Physics}\ }\textbf {\bibinfo {volume}
  {14}},\ \bibinfo {pages} {105003} (\bibinfo {year} {2012})}\BibitemShut
  {NoStop}%
\bibitem [{\citenamefont {Bajoni}(2012)}]{Bajoni2012}%
  \BibitemOpen
  \bibfield  {author} {\bibinfo {author} {\bibfnamefont {D.}~\bibnamefont
  {Bajoni}},\ }\href {http://stacks.iop.org/0022-3727/45/i=31/a=313001}
  {\bibfield  {journal} {\bibinfo  {journal} {Journal of Physics D: Applied
  Physics}\ }\textbf {\bibinfo {volume} {45}},\ \bibinfo {pages} {313001}
  (\bibinfo {year} {2012})}\BibitemShut {NoStop}%
\bibitem [{\citenamefont {Liew}\ \emph {et~al.}(2008)\citenamefont {Liew},
  \citenamefont {Kavokin},\ and\ \citenamefont {Shelykh}}]{Liew2008}%
  \BibitemOpen
  \bibfield  {author} {\bibinfo {author} {\bibfnamefont {T.~C.~H.}\
  \bibnamefont {Liew}}, \bibinfo {author} {\bibfnamefont {A.~V.}\ \bibnamefont
  {Kavokin}}, \ and\ \bibinfo {author} {\bibfnamefont {I.~A.}\ \bibnamefont
  {Shelykh}},\ }\href {\doibase 10.1103/PhysRevLett.101.016402} {\bibfield
  {journal} {\bibinfo  {journal} {Phys. Rev. Lett.}\ }\textbf {\bibinfo
  {volume} {101}},\ \bibinfo {pages} {016402} (\bibinfo {year}
  {2008})}\BibitemShut {NoStop}%
\bibitem [{\citenamefont {Espinosa-Ortega}\ and\ \citenamefont
  {Liew}(2013)}]{Ortega2013}%
  \BibitemOpen
  \bibfield  {author} {\bibinfo {author} {\bibfnamefont {T.}~\bibnamefont
  {Espinosa-Ortega}}\ and\ \bibinfo {author} {\bibfnamefont {T.~C.~H.}\
  \bibnamefont {Liew}},\ }\href {\doibase 10.1103/PhysRevB.87.195305}
  {\bibfield  {journal} {\bibinfo  {journal} {Phys. Rev. B}\ }\textbf {\bibinfo
  {volume} {87}},\ \bibinfo {pages} {195305} (\bibinfo {year}
  {2013})}\BibitemShut {NoStop}%
\bibitem [{\citenamefont {Sturm}\ \emph {et~al.}(2014)\citenamefont {Sturm},
  \citenamefont {Tanese}, \citenamefont {Nguyen}, \citenamefont {Flayac},
  \citenamefont {Galopin}, \citenamefont {Lema\^{i}tre}, \citenamefont
  {Sagnes}, \citenamefont {Solnyshkov}, \citenamefont {Amo}, \citenamefont
  {Malpuech},\ and\ \citenamefont {Bloch}}]{Sturm2014}%
  \BibitemOpen
  \bibfield  {author} {\bibinfo {author} {\bibfnamefont {C.}~\bibnamefont
  {Sturm}}, \bibinfo {author} {\bibfnamefont {D.}~\bibnamefont {Tanese}},
  \bibinfo {author} {\bibfnamefont {H.~S.}\ \bibnamefont {Nguyen}}, \bibinfo
  {author} {\bibfnamefont {H.}~\bibnamefont {Flayac}}, \bibinfo {author}
  {\bibfnamefont {E.}~\bibnamefont {Galopin}}, \bibinfo {author} {\bibfnamefont
  {A.}~\bibnamefont {Lema\^{i}tre}}, \bibinfo {author} {\bibfnamefont
  {I.}~\bibnamefont {Sagnes}}, \bibinfo {author} {\bibfnamefont
  {D.}~\bibnamefont {Solnyshkov}}, \bibinfo {author} {\bibfnamefont
  {A.}~\bibnamefont {Amo}}, \bibinfo {author} {\bibfnamefont {G.}~\bibnamefont
  {Malpuech}}, \ and\ \bibinfo {author} {\bibfnamefont {J.}~\bibnamefont
  {Bloch}},\ }\href {http://dx.doi.org/10.1038/ncomms4278} {\bibfield
  {journal} {\bibinfo  {journal} {Nat. Commun.}\ }\textbf {\bibinfo {volume}
  {5}},\ \bibinfo {pages} {3278} (\bibinfo {year} {2014})}\BibitemShut
  {NoStop}%
\bibitem [{\citenamefont {Kavokin}\ \emph {et~al.}(2012)\citenamefont
  {Kavokin}, \citenamefont {Shelykh}, \citenamefont {Taylor},\ and\
  \citenamefont {Glazov}}]{Kavokin2012}%
  \BibitemOpen
  \bibfield  {author} {\bibinfo {author} {\bibfnamefont {A.~V.}\ \bibnamefont
  {Kavokin}}, \bibinfo {author} {\bibfnamefont {I.~A.}\ \bibnamefont
  {Shelykh}}, \bibinfo {author} {\bibfnamefont {T.}~\bibnamefont {Taylor}}, \
  and\ \bibinfo {author} {\bibfnamefont {M.~M.}\ \bibnamefont {Glazov}},\
  }\href {\doibase 10.1103/PhysRevLett.108.197401} {\bibfield  {journal}
  {\bibinfo  {journal} {Phys. Rev. Lett.}\ }\textbf {\bibinfo {volume} {108}},\
  \bibinfo {pages} {197401} (\bibinfo {year} {2012})}\BibitemShut {NoStop}%
\bibitem [{\citenamefont {Fischer}\ \emph {et~al.}(2005)\citenamefont
  {Fischer}, \citenamefont {Hoffmann}, \citenamefont {Helm}, \citenamefont
  {Wilk}, \citenamefont {Rutz}, \citenamefont {Kleine-Ostmann}, \citenamefont
  {Koch},\ and\ \citenamefont {Jepsen}}]{Fischer2005}%
  \BibitemOpen
  \bibfield  {author} {\bibinfo {author} {\bibfnamefont {B.}~\bibnamefont
  {Fischer}}, \bibinfo {author} {\bibfnamefont {M.}~\bibnamefont {Hoffmann}},
  \bibinfo {author} {\bibfnamefont {H.}~\bibnamefont {Helm}}, \bibinfo {author}
  {\bibfnamefont {R.}~\bibnamefont {Wilk}}, \bibinfo {author} {\bibfnamefont
  {F.}~\bibnamefont {Rutz}}, \bibinfo {author} {\bibfnamefont {T.}~\bibnamefont
  {Kleine-Ostmann}}, \bibinfo {author} {\bibfnamefont {M.}~\bibnamefont
  {Koch}}, \ and\ \bibinfo {author} {\bibfnamefont {P.}~\bibnamefont
  {Jepsen}},\ }\href {\doibase 10.1364/OPEX.13.005205} {\bibfield  {journal}
  {\bibinfo  {journal} {Opt. Express}\ }\textbf {\bibinfo {volume} {13}},\
  \bibinfo {pages} {5205} (\bibinfo {year} {2005})}\BibitemShut {NoStop}%
\bibitem [{\citenamefont {Nagai}\ \emph {et~al.}(2004)\citenamefont {Nagai},
  \citenamefont {Imai}, \citenamefont {Fukasawa}, \citenamefont {Kato},\ and\
  \citenamefont {Yamauchi}}]{Nagai2004}%
  \BibitemOpen
  \bibfield  {author} {\bibinfo {author} {\bibfnamefont {N.}~\bibnamefont
  {Nagai}}, \bibinfo {author} {\bibfnamefont {T.}~\bibnamefont {Imai}},
  \bibinfo {author} {\bibfnamefont {R.}~\bibnamefont {Fukasawa}}, \bibinfo
  {author} {\bibfnamefont {K.}~\bibnamefont {Kato}}, \ and\ \bibinfo {author}
  {\bibfnamefont {K.}~\bibnamefont {Yamauchi}},\ }\href {\doibase
  http://dx.doi.org/10.1063/1.1811795} {\bibfield  {journal} {\bibinfo
  {journal} {Applied Physics Letters}\ }\textbf {\bibinfo {volume} {85}},\
  \bibinfo {pages} {4010} (\bibinfo {year} {2004})}\BibitemShut {NoStop}%
\bibitem [{\citenamefont {Kawase}\ \emph {et~al.}(2003)\citenamefont {Kawase},
  \citenamefont {Ogawa}, \citenamefont {Watanabe},\ and\ \citenamefont
  {Inoue}}]{Kawase2003}%
  \BibitemOpen
  \bibfield  {author} {\bibinfo {author} {\bibfnamefont {K.}~\bibnamefont
  {Kawase}}, \bibinfo {author} {\bibfnamefont {Y.}~\bibnamefont {Ogawa}},
  \bibinfo {author} {\bibfnamefont {Y.}~\bibnamefont {Watanabe}}, \ and\
  \bibinfo {author} {\bibfnamefont {H.}~\bibnamefont {Inoue}},\ }\href
  {\doibase 10.1364/OE.11.002549} {\bibfield  {journal} {\bibinfo  {journal}
  {Opt. Express}\ }\textbf {\bibinfo {volume} {11}},\ \bibinfo {pages} {2549}
  (\bibinfo {year} {2003})}\BibitemShut {NoStop}%
\bibitem [{\citenamefont {Tonouchi}(2007)}]{Tonouchi2007}%
  \BibitemOpen
  \bibfield  {author} {\bibinfo {author} {\bibfnamefont {M.}~\bibnamefont
  {Tonouchi}},\ }\href {http://dx.doi.org/10.1038/nphoton.2007.3} {\bibfield
  {journal} {\bibinfo  {journal} {Nat. Photon.}\ }\textbf {\bibinfo {volume}
  {1}},\ \bibinfo {pages} {97} (\bibinfo {year} {2007})}\BibitemShut {NoStop}%
\bibitem [{\citenamefont {Brown}\ \emph {et~al.}(1995)\citenamefont {Brown},
  \citenamefont {McIntosh}, \citenamefont {Nichols},\ and\ \citenamefont
  {Dennis}}]{Brown1995}%
  \BibitemOpen
  \bibfield  {author} {\bibinfo {author} {\bibfnamefont {E.~R.}\ \bibnamefont
  {Brown}}, \bibinfo {author} {\bibfnamefont {K.~A.}\ \bibnamefont {McIntosh}},
  \bibinfo {author} {\bibfnamefont {K.~B.}\ \bibnamefont {Nichols}}, \ and\
  \bibinfo {author} {\bibfnamefont {C.~L.}\ \bibnamefont {Dennis}},\ }\href
  {\doibase http://dx.doi.org/10.1063/1.113519} {\bibfield  {journal} {\bibinfo
   {journal} {Applied Physics Letters}\ }\textbf {\bibinfo {volume} {66}},\
  \bibinfo {pages} {285} (\bibinfo {year} {1995})}\BibitemShut {NoStop}%
\bibitem [{\citenamefont {Cai}\ \emph {et~al.}(1997)\citenamefont {Cai},
  \citenamefont {Brener}, \citenamefont {Lopata}, \citenamefont {Wynn},
  \citenamefont {Pfeiffer},\ and\ \citenamefont {Federici}}]{Cai1997}%
  \BibitemOpen
  \bibfield  {author} {\bibinfo {author} {\bibfnamefont {Y.}~\bibnamefont
  {Cai}}, \bibinfo {author} {\bibfnamefont {I.}~\bibnamefont {Brener}},
  \bibinfo {author} {\bibfnamefont {J.}~\bibnamefont {Lopata}}, \bibinfo
  {author} {\bibfnamefont {J.}~\bibnamefont {Wynn}}, \bibinfo {author}
  {\bibfnamefont {L.}~\bibnamefont {Pfeiffer}}, \ and\ \bibinfo {author}
  {\bibfnamefont {J.}~\bibnamefont {Federici}},\ }\href {\doibase
  http://dx.doi.org/10.1063/1.119346} {\bibfield  {journal} {\bibinfo
  {journal} {Applied Physics Letters}\ }\textbf {\bibinfo {volume} {71}},\
  \bibinfo {pages} {2076} (\bibinfo {year} {1997})}\BibitemShut {NoStop}%
\bibitem [{\citenamefont {Williams}(2007)}]{Williams2007}%
  \BibitemOpen
  \bibfield  {author} {\bibinfo {author} {\bibfnamefont {B.~S.}\ \bibnamefont
  {Williams}},\ }\href {http://dx.doi.org/10.1038/nphoton.2007.166} {\bibfield
  {journal} {\bibinfo  {journal} {Nat. Photon.}\ }\textbf {\bibinfo {volume}
  {1}},\ \bibinfo {pages} {517} (\bibinfo {year} {2007})}\BibitemShut {NoStop}%
\bibitem [{\citenamefont {Mittleman}(2013)}]{Mittleman2013}%
  \BibitemOpen
  \bibfield  {author} {\bibinfo {author} {\bibfnamefont {D.~M.}\ \bibnamefont
  {Mittleman}},\ }\href {http://dx.doi.org/10.1038/nphoton.2013.235} {\bibfield
   {journal} {\bibinfo  {journal} {Nat. Photon.}\ }\textbf {\bibinfo {volume}
  {7}},\ \bibinfo {pages} {666} (\bibinfo {year} {2013})}\BibitemShut {NoStop}%
\bibitem [{\citenamefont {Schneider}\ \emph {et~al.}(2013)\citenamefont
  {Schneider}, \citenamefont {Rahimi-Iman}, \citenamefont {Kim}, \citenamefont
  {Fischer}, \citenamefont {Savenko}, \citenamefont {Amthor}, \citenamefont
  {Lermer}, \citenamefont {Wolf}, \citenamefont {Worschech}, \citenamefont
  {Kulakovskii}, \citenamefont {Shelykh}, \citenamefont {Kamp}, \citenamefont
  {Reitzenstein}, \citenamefont {Forchel}, \citenamefont {Yamamoto},\ and\
  \citenamefont {H\"ofling}}]{Schneider2013}%
  \BibitemOpen
  \bibfield  {author} {\bibinfo {author} {\bibfnamefont {C.}~\bibnamefont
  {Schneider}}, \bibinfo {author} {\bibfnamefont {A.}~\bibnamefont
  {Rahimi-Iman}}, \bibinfo {author} {\bibfnamefont {N.~Y.}\ \bibnamefont
  {Kim}}, \bibinfo {author} {\bibfnamefont {J.}~\bibnamefont {Fischer}},
  \bibinfo {author} {\bibfnamefont {I.~G.}\ \bibnamefont {Savenko}}, \bibinfo
  {author} {\bibfnamefont {M.}~\bibnamefont {Amthor}}, \bibinfo {author}
  {\bibfnamefont {M.}~\bibnamefont {Lermer}}, \bibinfo {author} {\bibfnamefont
  {A.}~\bibnamefont {Wolf}}, \bibinfo {author} {\bibfnamefont {L.}~\bibnamefont
  {Worschech}}, \bibinfo {author} {\bibfnamefont {V.~D.}\ \bibnamefont
  {Kulakovskii}}, \bibinfo {author} {\bibfnamefont {I.~A.}\ \bibnamefont
  {Shelykh}}, \bibinfo {author} {\bibfnamefont {M.}~\bibnamefont {Kamp}},
  \bibinfo {author} {\bibfnamefont {S.}~\bibnamefont {Reitzenstein}}, \bibinfo
  {author} {\bibfnamefont {A.}~\bibnamefont {Forchel}}, \bibinfo {author}
  {\bibfnamefont {Y.}~\bibnamefont {Yamamoto}}, \ and\ \bibinfo {author}
  {\bibfnamefont {S.}~\bibnamefont {H\"ofling}},\ }\href
  {http://dx.doi.org/10.1038/nature12036} {\bibfield  {journal} {\bibinfo
  {journal} {Nature}\ }\textbf {\bibinfo {volume} {497}},\ \bibinfo {pages}
  {348} (\bibinfo {year} {2013})}\BibitemShut {NoStop}%
\bibitem [{\citenamefont {Bhattacharya}\ \emph {et~al.}(2013)\citenamefont
  {Bhattacharya}, \citenamefont {Xiao}, \citenamefont {Das}, \citenamefont
  {Bhowmick},\ and\ \citenamefont {Heo}}]{Bhattacharya2013}%
  \BibitemOpen
  \bibfield  {author} {\bibinfo {author} {\bibfnamefont {P.}~\bibnamefont
  {Bhattacharya}}, \bibinfo {author} {\bibfnamefont {B.}~\bibnamefont {Xiao}},
  \bibinfo {author} {\bibfnamefont {A.}~\bibnamefont {Das}}, \bibinfo {author}
  {\bibfnamefont {S.}~\bibnamefont {Bhowmick}}, \ and\ \bibinfo {author}
  {\bibfnamefont {J.}~\bibnamefont {Heo}},\ }\href {\doibase
  10.1103/PhysRevLett.110.206403} {\bibfield  {journal} {\bibinfo  {journal}
  {Phys. Rev. Lett.}\ }\textbf {\bibinfo {volume} {110}},\ \bibinfo {pages}
  {206403} (\bibinfo {year} {2013})}\BibitemShut {NoStop}%
\bibitem [{\citenamefont {Bhattacharya}\ \emph {et~al.}(2014)\citenamefont
  {Bhattacharya}, \citenamefont {Frost}, \citenamefont {Deshpande},
  \citenamefont {Baten}, \citenamefont {Hazari},\ and\ \citenamefont
  {Das}}]{Bhattacharya2014}%
  \BibitemOpen
  \bibfield  {author} {\bibinfo {author} {\bibfnamefont {P.}~\bibnamefont
  {Bhattacharya}}, \bibinfo {author} {\bibfnamefont {T.}~\bibnamefont {Frost}},
  \bibinfo {author} {\bibfnamefont {S.}~\bibnamefont {Deshpande}}, \bibinfo
  {author} {\bibfnamefont {M.~Z.}\ \bibnamefont {Baten}}, \bibinfo {author}
  {\bibfnamefont {A.}~\bibnamefont {Hazari}}, \ and\ \bibinfo {author}
  {\bibfnamefont {A.}~\bibnamefont {Das}},\ }\href {\doibase
  10.1103/PhysRevLett.112.236802} {\bibfield  {journal} {\bibinfo  {journal}
  {Phys. Rev. Lett.}\ }\textbf {\bibinfo {volume} {112}},\ \bibinfo {pages}
  {236802} (\bibinfo {year} {2014})}\BibitemShut {NoStop}%
\bibitem [{\citenamefont {Tsintzos}\ \emph {et~al.}(2008)\citenamefont
  {Tsintzos}, \citenamefont {Pelekanos}, \citenamefont {Konstantinidis},
  \citenamefont {Hatzopoulos},\ and\ \citenamefont {Savvidis}}]{Tsintzos2008}%
  \BibitemOpen
  \bibfield  {author} {\bibinfo {author} {\bibfnamefont {S.~I.}\ \bibnamefont
  {Tsintzos}}, \bibinfo {author} {\bibfnamefont {N.~T.}\ \bibnamefont
  {Pelekanos}}, \bibinfo {author} {\bibfnamefont {G.}~\bibnamefont
  {Konstantinidis}}, \bibinfo {author} {\bibfnamefont {Z.}~\bibnamefont
  {Hatzopoulos}}, \ and\ \bibinfo {author} {\bibfnamefont {P.~G.}\ \bibnamefont
  {Savvidis}},\ }\href {http://dx.doi.org/10.1038/nature06979} {\bibfield
  {journal} {\bibinfo  {journal} {Nature}\ }\textbf {\bibinfo {volume} {453}},\
  \bibinfo {pages} {372} (\bibinfo {year} {2008})}\BibitemShut {NoStop}%
\bibitem [{\citenamefont {Ballarini}\ \emph {et~al.}(2013)\citenamefont
  {Ballarini}, \citenamefont {De~Giorgi}, \citenamefont {Cancellieri},
  \citenamefont {Houdr\"e}, \citenamefont {Giacobino}, \citenamefont
  {Cingolani}, \citenamefont {Bramati}, \citenamefont {Gigli},\ and\
  \citenamefont {Sanvitto}}]{Ballarini2013}%
  \BibitemOpen
  \bibfield  {author} {\bibinfo {author} {\bibfnamefont {D.}~\bibnamefont
  {Ballarini}}, \bibinfo {author} {\bibfnamefont {M.}~\bibnamefont
  {De~Giorgi}}, \bibinfo {author} {\bibfnamefont {E.}~\bibnamefont
  {Cancellieri}}, \bibinfo {author} {\bibfnamefont {R.}~\bibnamefont
  {Houdr\"e}}, \bibinfo {author} {\bibfnamefont {E.}~\bibnamefont {Giacobino}},
  \bibinfo {author} {\bibfnamefont {R.}~\bibnamefont {Cingolani}}, \bibinfo
  {author} {\bibfnamefont {A.}~\bibnamefont {Bramati}}, \bibinfo {author}
  {\bibfnamefont {G.}~\bibnamefont {Gigli}}, \ and\ \bibinfo {author}
  {\bibfnamefont {D.}~\bibnamefont {Sanvitto}},\ }\href
  {http://dx.doi.org/10.1038/ncomms2734} {\bibfield  {journal} {\bibinfo
  {journal} {Nat. Commun.}\ }\textbf {\bibinfo {volume} {4}},\ \bibinfo {pages}
  {1778} (\bibinfo {year} {2013})}\BibitemShut {NoStop}%
\bibitem [{\citenamefont {Nguyen}\ \emph {et~al.}(2013)\citenamefont {Nguyen},
  \citenamefont {Vishnevsky}, \citenamefont {Sturm}, \citenamefont {Tanese},
  \citenamefont {Solnyshkov}, \citenamefont {Galopin}, \citenamefont
  {Lema\^{i}tre}, \citenamefont {Sagnes}, \citenamefont {Amo}, \citenamefont
  {Malpuech},\ and\ \citenamefont {Bloch}}]{Nguyen2013}%
  \BibitemOpen
  \bibfield  {author} {\bibinfo {author} {\bibfnamefont {H.~S.}\ \bibnamefont
  {Nguyen}}, \bibinfo {author} {\bibfnamefont {D.}~\bibnamefont {Vishnevsky}},
  \bibinfo {author} {\bibfnamefont {C.}~\bibnamefont {Sturm}}, \bibinfo
  {author} {\bibfnamefont {D.}~\bibnamefont {Tanese}}, \bibinfo {author}
  {\bibfnamefont {D.}~\bibnamefont {Solnyshkov}}, \bibinfo {author}
  {\bibfnamefont {E.}~\bibnamefont {Galopin}}, \bibinfo {author} {\bibfnamefont
  {A.}~\bibnamefont {Lema\^{i}tre}}, \bibinfo {author} {\bibfnamefont
  {I.}~\bibnamefont {Sagnes}}, \bibinfo {author} {\bibfnamefont
  {A.}~\bibnamefont {Amo}}, \bibinfo {author} {\bibfnamefont {G.}~\bibnamefont
  {Malpuech}}, \ and\ \bibinfo {author} {\bibfnamefont {J.}~\bibnamefont
  {Bloch}},\ }\href {\doibase 10.1103/PhysRevLett.110.236601} {\bibfield
  {journal} {\bibinfo  {journal} {Phys. Rev. Lett.}\ }\textbf {\bibinfo
  {volume} {110}},\ \bibinfo {pages} {236601} (\bibinfo {year}
  {2013})}\BibitemShut {NoStop}%
\bibitem [{\citenamefont {Gao}\ \emph {et~al.}(2012)\citenamefont {Gao},
  \citenamefont {Eldridge}, \citenamefont {Liew}, \citenamefont {Tsintzos},
  \citenamefont {Stavrinidis}, \citenamefont {Deligeorgis}, \citenamefont
  {Hatzopoulos},\ and\ \citenamefont {Savvidis}}]{Gao2012}%
  \BibitemOpen
  \bibfield  {author} {\bibinfo {author} {\bibfnamefont {T.}~\bibnamefont
  {Gao}}, \bibinfo {author} {\bibfnamefont {P.~S.}\ \bibnamefont {Eldridge}},
  \bibinfo {author} {\bibfnamefont {T.~C.~H.}\ \bibnamefont {Liew}}, \bibinfo
  {author} {\bibfnamefont {S.~I.}\ \bibnamefont {Tsintzos}}, \bibinfo {author}
  {\bibfnamefont {G.}~\bibnamefont {Stavrinidis}}, \bibinfo {author}
  {\bibfnamefont {G.}~\bibnamefont {Deligeorgis}}, \bibinfo {author}
  {\bibfnamefont {Z.}~\bibnamefont {Hatzopoulos}}, \ and\ \bibinfo {author}
  {\bibfnamefont {P.~G.}\ \bibnamefont {Savvidis}},\ }\href {\doibase
  10.1103/PhysRevB.85.235102} {\bibfield  {journal} {\bibinfo  {journal} {Phys.
  Rev. B}\ }\textbf {\bibinfo {volume} {85}},\ \bibinfo {pages} {235102}
  (\bibinfo {year} {2012})}\BibitemShut {NoStop}%
\bibitem [{\citenamefont {Hayat}\ \emph {et~al.}(2012)\citenamefont {Hayat},
  \citenamefont {Lange}, \citenamefont {Rozema}, \citenamefont {Darabi},
  \citenamefont {van Driel}, \citenamefont {Steinberg}, \citenamefont {Nelsen},
  \citenamefont {Snoke}, \citenamefont {Pfeiffer},\ and\ \citenamefont
  {West}}]{Hayat2012}%
  \BibitemOpen
  \bibfield  {author} {\bibinfo {author} {\bibfnamefont {A.}~\bibnamefont
  {Hayat}}, \bibinfo {author} {\bibfnamefont {C.}~\bibnamefont {Lange}},
  \bibinfo {author} {\bibfnamefont {L.~A.}\ \bibnamefont {Rozema}}, \bibinfo
  {author} {\bibfnamefont {A.}~\bibnamefont {Darabi}}, \bibinfo {author}
  {\bibfnamefont {H.~M.}\ \bibnamefont {van Driel}}, \bibinfo {author}
  {\bibfnamefont {A.~M.}\ \bibnamefont {Steinberg}}, \bibinfo {author}
  {\bibfnamefont {B.}~\bibnamefont {Nelsen}}, \bibinfo {author} {\bibfnamefont
  {D.~W.}\ \bibnamefont {Snoke}}, \bibinfo {author} {\bibfnamefont {L.~N.}\
  \bibnamefont {Pfeiffer}}, \ and\ \bibinfo {author} {\bibfnamefont {K.~W.}\
  \bibnamefont {West}},\ }\href {\doibase 10.1103/PhysRevLett.109.033605}
  {\bibfield  {journal} {\bibinfo  {journal} {Phys. Rev. Lett.}\ }\textbf
  {\bibinfo {volume} {109}},\ \bibinfo {pages} {033605} (\bibinfo {year}
  {2012})}\BibitemShut {NoStop}%
\bibitem [{\citenamefont {Cancellieri}\ \emph {et~al.}(2014)\citenamefont
  {Cancellieri}, \citenamefont {Hayat}, \citenamefont {Steinberg},
  \citenamefont {Giacobino},\ and\ \citenamefont {Bramati}}]{Cancellieri2014}%
  \BibitemOpen
  \bibfield  {author} {\bibinfo {author} {\bibfnamefont {E.}~\bibnamefont
  {Cancellieri}}, \bibinfo {author} {\bibfnamefont {A.}~\bibnamefont {Hayat}},
  \bibinfo {author} {\bibfnamefont {A.~M.}\ \bibnamefont {Steinberg}}, \bibinfo
  {author} {\bibfnamefont {E.}~\bibnamefont {Giacobino}}, \ and\ \bibinfo
  {author} {\bibfnamefont {A.}~\bibnamefont {Bramati}},\ }\href {\doibase
  10.1103/PhysRevLett.112.053601} {\bibfield  {journal} {\bibinfo  {journal}
  {Phys. Rev. Lett.}\ }\textbf {\bibinfo {volume} {112}},\ \bibinfo {pages}
  {053601} (\bibinfo {year} {2014})}\BibitemShut {NoStop}%
\bibitem [{\citenamefont {De~Giorgi}\ \emph {et~al.}(2014)\citenamefont
  {De~Giorgi}, \citenamefont {Ballarini}, \citenamefont {Cazzato},
  \citenamefont {Deligeorgis}, \citenamefont {Tsintzos}, \citenamefont
  {Hatzopoulos}, \citenamefont {Savvidis}, \citenamefont {Gigli}, \citenamefont
  {Laussy},\ and\ \citenamefont {Sanvitto}}]{DeGiorgi2014}%
  \BibitemOpen
  \bibfield  {author} {\bibinfo {author} {\bibfnamefont {M.}~\bibnamefont
  {De~Giorgi}}, \bibinfo {author} {\bibfnamefont {D.}~\bibnamefont
  {Ballarini}}, \bibinfo {author} {\bibfnamefont {P.}~\bibnamefont {Cazzato}},
  \bibinfo {author} {\bibfnamefont {G.}~\bibnamefont {Deligeorgis}}, \bibinfo
  {author} {\bibfnamefont {S.~I.}\ \bibnamefont {Tsintzos}}, \bibinfo {author}
  {\bibfnamefont {Z.}~\bibnamefont {Hatzopoulos}}, \bibinfo {author}
  {\bibfnamefont {P.~G.}\ \bibnamefont {Savvidis}}, \bibinfo {author}
  {\bibfnamefont {G.}~\bibnamefont {Gigli}}, \bibinfo {author} {\bibfnamefont
  {F.~P.}\ \bibnamefont {Laussy}}, \ and\ \bibinfo {author} {\bibfnamefont
  {D.}~\bibnamefont {Sanvitto}},\ }\href {\doibase
  10.1103/PhysRevLett.112.113602} {\bibfield  {journal} {\bibinfo  {journal}
  {Phys. Rev. Lett.}\ }\textbf {\bibinfo {volume} {112}},\ \bibinfo {pages}
  {113602} (\bibinfo {year} {2014})}\BibitemShut {NoStop}%
\bibitem [{\citenamefont {Lem\'{e}nager}\ \emph {et~al.}(2014)\citenamefont
  {Lem\'{e}nager}, \citenamefont {Pisanello}, \citenamefont {Bloch},
  \citenamefont {Kavokin}, \citenamefont {Amo}, \citenamefont {Lemaitre},
  \citenamefont {Galopin}, \citenamefont {Sagnes}, \citenamefont {Vittorio},
  \citenamefont {Giacobino},\ and\ \citenamefont {Bramati}}]{Lemenager2014}%
  \BibitemOpen
  \bibfield  {author} {\bibinfo {author} {\bibfnamefont {G.}~\bibnamefont
  {Lem\'{e}nager}}, \bibinfo {author} {\bibfnamefont {F.}~\bibnamefont
  {Pisanello}}, \bibinfo {author} {\bibfnamefont {J.}~\bibnamefont {Bloch}},
  \bibinfo {author} {\bibfnamefont {A.}~\bibnamefont {Kavokin}}, \bibinfo
  {author} {\bibfnamefont {A.}~\bibnamefont {Amo}}, \bibinfo {author}
  {\bibfnamefont {A.}~\bibnamefont {Lemaitre}}, \bibinfo {author}
  {\bibfnamefont {E.}~\bibnamefont {Galopin}}, \bibinfo {author} {\bibfnamefont
  {I.}~\bibnamefont {Sagnes}}, \bibinfo {author} {\bibfnamefont {M.~D.}\
  \bibnamefont {Vittorio}}, \bibinfo {author} {\bibfnamefont {E.}~\bibnamefont
  {Giacobino}}, \ and\ \bibinfo {author} {\bibfnamefont {A.}~\bibnamefont
  {Bramati}},\ }\href {\doibase 10.1364/OL.39.000307} {\bibfield  {journal}
  {\bibinfo  {journal} {Opt. Lett.}\ }\textbf {\bibinfo {volume} {39}},\
  \bibinfo {pages} {307} (\bibinfo {year} {2014})}\BibitemShut {NoStop}%
\bibitem [{\citenamefont {Schmutzler}\ \emph {et~al.}(2013)\citenamefont
  {Schmutzler}, \citenamefont {Veit}, \citenamefont {A\ss{}mann}, \citenamefont
  {Tempel}, \citenamefont {H\"ofling}, \citenamefont {Kamp}, \citenamefont
  {Forchel},\ and\ \citenamefont {Bayer}}]{Schmutzler2013a}%
  \BibitemOpen
  \bibfield  {author} {\bibinfo {author} {\bibfnamefont {J.}~\bibnamefont
  {Schmutzler}}, \bibinfo {author} {\bibfnamefont {F.}~\bibnamefont {Veit}},
  \bibinfo {author} {\bibfnamefont {M.}~\bibnamefont {A\ss{}mann}}, \bibinfo
  {author} {\bibfnamefont {J.-S.}\ \bibnamefont {Tempel}}, \bibinfo {author}
  {\bibfnamefont {S.}~\bibnamefont {H\"ofling}}, \bibinfo {author}
  {\bibfnamefont {M.}~\bibnamefont {Kamp}}, \bibinfo {author} {\bibfnamefont
  {A.}~\bibnamefont {Forchel}}, \ and\ \bibinfo {author} {\bibfnamefont
  {M.}~\bibnamefont {Bayer}},\ }\href {\doibase 10.1063/1.4794144} {\bibfield
  {journal} {\bibinfo  {journal} {Applied Physics Letters}\ }\textbf {\bibinfo
  {volume} {102}},\ \bibinfo {eid} {081115} (\bibinfo {year}
  {2013})}\BibitemShut {NoStop}%
\bibitem [{\citenamefont {Kavokin}\ \emph {et~al.}(2010)\citenamefont
  {Kavokin}, \citenamefont {Kaliteevski}, \citenamefont {Abram}, \citenamefont
  {Kavokin}, \citenamefont {Sharkova},\ and\ \citenamefont
  {Shelykh}}]{Kavokin2010}%
  \BibitemOpen
  \bibfield  {author} {\bibinfo {author} {\bibfnamefont {K.~V.}\ \bibnamefont
  {Kavokin}}, \bibinfo {author} {\bibfnamefont {M.~A.}\ \bibnamefont
  {Kaliteevski}}, \bibinfo {author} {\bibfnamefont {R.~A.}\ \bibnamefont
  {Abram}}, \bibinfo {author} {\bibfnamefont {A.~V.}\ \bibnamefont {Kavokin}},
  \bibinfo {author} {\bibfnamefont {S.}~\bibnamefont {Sharkova}}, \ and\
  \bibinfo {author} {\bibfnamefont {I.~A.}\ \bibnamefont {Shelykh}},\ }\href
  {\doibase http://dx.doi.org/10.1063/1.3519978} {\bibfield  {journal}
  {\bibinfo  {journal} {Applied Physics Letters}\ }\textbf {\bibinfo {volume}
  {97}},\ \bibinfo {eid} {201111} (\bibinfo {year} {2010})}\BibitemShut
  {NoStop}%
\bibitem [{\citenamefont {Nithisoontorn}\ \emph {et~al.}(1989)\citenamefont
  {Nithisoontorn}, \citenamefont {Unterrainer}, \citenamefont {Michaelis},
  \citenamefont {Sawaki}, \citenamefont {Gornik},\ and\ \citenamefont
  {Kano}}]{Nithisoontorn1989}%
  \BibitemOpen
  \bibfield  {author} {\bibinfo {author} {\bibfnamefont {M.}~\bibnamefont
  {Nithisoontorn}}, \bibinfo {author} {\bibfnamefont {K.}~\bibnamefont
  {Unterrainer}}, \bibinfo {author} {\bibfnamefont {S.}~\bibnamefont
  {Michaelis}}, \bibinfo {author} {\bibfnamefont {N.}~\bibnamefont {Sawaki}},
  \bibinfo {author} {\bibfnamefont {E.}~\bibnamefont {Gornik}}, \ and\ \bibinfo
  {author} {\bibfnamefont {H.}~\bibnamefont {Kano}},\ }\href {\doibase
  10.1103/PhysRevLett.62.3078} {\bibfield  {journal} {\bibinfo  {journal}
  {Phys. Rev. Lett.}\ }\textbf {\bibinfo {volume} {62}},\ \bibinfo {pages}
  {3078} (\bibinfo {year} {1989})}\BibitemShut {NoStop}%
\bibitem [{\citenamefont {Catalano}\ \emph {et~al.}(1989)\citenamefont
  {Catalano}, \citenamefont {Cingolani}, \citenamefont {Cingolani},
  \citenamefont {Lepore},\ and\ \citenamefont {Ploog}}]{Catalano1989}%
  \BibitemOpen
  \bibfield  {author} {\bibinfo {author} {\bibfnamefont {I.~M.}\ \bibnamefont
  {Catalano}}, \bibinfo {author} {\bibfnamefont {A.}~\bibnamefont {Cingolani}},
  \bibinfo {author} {\bibfnamefont {R.}~\bibnamefont {Cingolani}}, \bibinfo
  {author} {\bibfnamefont {M.}~\bibnamefont {Lepore}}, \ and\ \bibinfo {author}
  {\bibfnamefont {K.}~\bibnamefont {Ploog}},\ }\href {\doibase
  10.1103/PhysRevB.40.1312} {\bibfield  {journal} {\bibinfo  {journal} {Phys.
  Rev. B}\ }\textbf {\bibinfo {volume} {40}},\ \bibinfo {pages} {1312}
  (\bibinfo {year} {1989})}\BibitemShut {NoStop}%
\bibitem [{\citenamefont {Catalano}\ \emph {et~al.}(1990)\citenamefont
  {Catalano}, \citenamefont {Cingolani}, \citenamefont {Lepore}, \citenamefont
  {Cingolani},\ and\ \citenamefont {Ploog}}]{Catalano1990}%
  \BibitemOpen
  \bibfield  {author} {\bibinfo {author} {\bibfnamefont {I.~M.}\ \bibnamefont
  {Catalano}}, \bibinfo {author} {\bibfnamefont {A.}~\bibnamefont {Cingolani}},
  \bibinfo {author} {\bibfnamefont {M.}~\bibnamefont {Lepore}}, \bibinfo
  {author} {\bibfnamefont {R.}~\bibnamefont {Cingolani}}, \ and\ \bibinfo
  {author} {\bibfnamefont {K.}~\bibnamefont {Ploog}},\ }\href {\doibase
  10.1103/PhysRevB.41.12937} {\bibfield  {journal} {\bibinfo  {journal} {Phys.
  Rev. B}\ }\textbf {\bibinfo {volume} {41}},\ \bibinfo {pages} {12937}
  (\bibinfo {year} {1990})}\BibitemShut {NoStop}%
\bibitem [{\citenamefont {Tai}\ \emph {et~al.}(1989)\citenamefont {Tai},
  \citenamefont {Mysyrowicz}, \citenamefont {Fischer}, \citenamefont
  {Slusher},\ and\ \citenamefont {Cho}}]{Tai1989}%
  \BibitemOpen
  \bibfield  {author} {\bibinfo {author} {\bibfnamefont {K.}~\bibnamefont
  {Tai}}, \bibinfo {author} {\bibfnamefont {A.}~\bibnamefont {Mysyrowicz}},
  \bibinfo {author} {\bibfnamefont {R.~J.}\ \bibnamefont {Fischer}}, \bibinfo
  {author} {\bibfnamefont {R.~E.}\ \bibnamefont {Slusher}}, \ and\ \bibinfo
  {author} {\bibfnamefont {A.~Y.}\ \bibnamefont {Cho}},\ }\href {\doibase
  10.1103/PhysRevLett.62.1784} {\bibfield  {journal} {\bibinfo  {journal}
  {Phys. Rev. Lett.}\ }\textbf {\bibinfo {volume} {62}},\ \bibinfo {pages}
  {1784} (\bibinfo {year} {1989})}\BibitemShut {NoStop}%
\bibitem [{\citenamefont {Chen}\ \emph {et~al.}(1997)\citenamefont {Chen},
  \citenamefont {Blake}, \citenamefont {Gaidis}, \citenamefont {Brown},
  \citenamefont {McIntosh}, \citenamefont {Chou}, \citenamefont {Nathan},\ and\
  \citenamefont {Williamson}}]{Chen1997}%
  \BibitemOpen
  \bibfield  {author} {\bibinfo {author} {\bibfnamefont {P.}~\bibnamefont
  {Chen}}, \bibinfo {author} {\bibfnamefont {G.~A.}\ \bibnamefont {Blake}},
  \bibinfo {author} {\bibfnamefont {M.~C.}\ \bibnamefont {Gaidis}}, \bibinfo
  {author} {\bibfnamefont {E.~R.}\ \bibnamefont {Brown}}, \bibinfo {author}
  {\bibfnamefont {K.~A.}\ \bibnamefont {McIntosh}}, \bibinfo {author}
  {\bibfnamefont {S.~Y.}\ \bibnamefont {Chou}}, \bibinfo {author}
  {\bibfnamefont {M.~I.}\ \bibnamefont {Nathan}}, \ and\ \bibinfo {author}
  {\bibfnamefont {F.}~\bibnamefont {Williamson}},\ }\href {\doibase
  http://dx.doi.org/10.1063/1.119845} {\bibfield  {journal} {\bibinfo
  {journal} {Applied Physics Letters}\ }\textbf {\bibinfo {volume} {71}},\
  \bibinfo {pages} {1601} (\bibinfo {year} {1997})}\BibitemShut {NoStop}%
\bibitem [{\citenamefont {Bloch}\ \emph {et~al.}(2005)\citenamefont {Bloch},
  \citenamefont {Sermage}, \citenamefont {Perrin}, \citenamefont {Senellart},
  \citenamefont {Andr\'e},\ and\ \citenamefont {Dang}}]{Bloch2005}%
  \BibitemOpen
  \bibfield  {author} {\bibinfo {author} {\bibfnamefont {J.}~\bibnamefont
  {Bloch}}, \bibinfo {author} {\bibfnamefont {B.}~\bibnamefont {Sermage}},
  \bibinfo {author} {\bibfnamefont {M.}~\bibnamefont {Perrin}}, \bibinfo
  {author} {\bibfnamefont {P.}~\bibnamefont {Senellart}}, \bibinfo {author}
  {\bibfnamefont {R.}~\bibnamefont {Andr\'e}}, \ and\ \bibinfo {author}
  {\bibfnamefont {L.~S.}\ \bibnamefont {Dang}},\ }\href {\doibase
  10.1103/PhysRevB.71.155311} {\bibfield  {journal} {\bibinfo  {journal} {Phys.
  Rev. B}\ }\textbf {\bibinfo {volume} {71}},\ \bibinfo {pages} {155311}
  (\bibinfo {year} {2005})}\BibitemShut {NoStop}%
\bibitem [{\citenamefont {Pavlov}\ \emph {et~al.}(2005)\citenamefont {Pavlov},
  \citenamefont {Kalashnikova}, \citenamefont {Pisarev}, \citenamefont
  {S\"anger}, \citenamefont {Yakovlev},\ and\ \citenamefont
  {Bayer}}]{Pavlov2005}%
  \BibitemOpen
  \bibfield  {author} {\bibinfo {author} {\bibfnamefont {V.~V.}\ \bibnamefont
  {Pavlov}}, \bibinfo {author} {\bibfnamefont {A.~M.}\ \bibnamefont
  {Kalashnikova}}, \bibinfo {author} {\bibfnamefont {R.~V.}\ \bibnamefont
  {Pisarev}}, \bibinfo {author} {\bibfnamefont {I.}~\bibnamefont {S\"anger}},
  \bibinfo {author} {\bibfnamefont {D.~R.}\ \bibnamefont {Yakovlev}}, \ and\
  \bibinfo {author} {\bibfnamefont {M.}~\bibnamefont {Bayer}},\ }\href
  {\doibase 10.1103/PhysRevLett.94.157404} {\bibfield  {journal} {\bibinfo
  {journal} {Phys. Rev. Lett.}\ }\textbf {\bibinfo {volume} {94}},\ \bibinfo
  {pages} {157404} (\bibinfo {year} {2005})}\BibitemShut {NoStop}%
\end{thebibliography}

%

\end{document}